\documentclass[conference£¬a4paper]{IEEEtran}

\usepackage{amsfonts}
\usepackage{amssymb}
\usepackage{stfloats}
\usepackage{cite}
\usepackage{graphicx}
\usepackage{epstopdf}
\usepackage{psfrag}
\usepackage{subfigure}
\usepackage{amsmath}
\usepackage{array}
\usepackage{stfloats}
\usepackage{multirow}
\usepackage{color}
\usepackage{booktabs}
\usepackage{url}
\usepackage{graphicx}

\title{Performance Evaluation for LTE-V based Vehicle-to-Vehicle Platooning Communication}

\author{Tao Yu, Shunqing Zhang,~\IEEEmembership{Senior Member, IEEE}, Shan Cao, 
and Shugong Xu,~\IEEEmembership{Fellow, IEEE}\\
\IEEEauthorblockA{Shanghai Institute for Advanced Communication and Data Science\\
Key laboratory of Specialty Fiber Optics and Optical Access Networks\\
Joint International Research Laboratory of Specialty Fiber Optics and Advanced Communication\\
Shanghai University, Shanghai 200444, China\\
E-mail: \{yutao\_burton, shunqing, cshan, shugong\}@shu.edu.cn}
}

\begin{document}
\maketitle

\begin{abstract}
With the raising demand for autonomous driving, vehicle-to-vehicle communications becomes a key technology enabler for the future intelligent transportation system. Based on our current knowledge field, there is limited network simulator that can support end-to-end performance evaluation for LTE-V based vehicle-to-vehicle platooning systems. To address this problem, we start with an integrated platform that combines traffic generator and network simulator together, and build the V2V transmission capability according to LTE-V specification. On top of that, we simulate the end-to-end throughput and delay profiles in different layers to compare different configurations of platooning systems. Through numerical experiments, we show that the LTE-V system is unable to support the highest degree of automation under shadowing effects in the vehicle platooning scenarios, which requires ultra-reliable low-latency communication enhancement in 5G networks. Meanwhile, the throughput and delay performance for vehicle platooning changes dramatically in PDCP layers, where we believe further improvements are necessary. 
\end{abstract}


\section{Introduction} \label{sect:intro}

With a rising demand for autonomous driving, the surrounding traffic environment sensing and intercommunication become one of the key design challenges in the intelligent transportation system (ITS). Vehicular networks \cite{karagiannis2011vehicular}, aiming to provide wireless information exchange channels for vehicle to anything (V2X) communication, are currently attractive research areas. Compared with the traditional dedicated short range communications (DSRC) solutions, cellular-based V2X communication provides significant improvement in quality-of-service (QoS) guarantee \cite{wang2017comparison} \cite{cecchini2017performance}  and deployment cost reduction \cite{chen2016lte}, and has recently been adopted as part of the 3rd Generation Partner Project (3GPP) Long-Term Evolution (LTE) Release 14 (R14) \cite{3gpp:v2x} (also known as LTE-V) and shown to be an important role in the future 5G networks \cite{molina2017lte} \cite{chen2017vehicle}. 

Based on LTE-V systems, vehicle-to-vehicle (V2V) communications, also known as mode 4 transmission in the 3GPP definition \cite{3gpp:frame}, are actively investigated in the current literature. For example, the wireless transmission resource allocation schemes have been proposed in \cite{wei2018wireless,ye2017deep,mei2018latency}, where the spectrum sharing, the wireless link assignment and the corresponding power adjustment are the key design factors to guarantee the transmission reliability and latency. In \cite{jeyaraj2017reliability}, the theoretical analysis on orthogonal street systems is reported. Although the above results provide a comprehensive study on the performance optimization in the physical and medium access layers, the following issues have {\em not} yet been investigated based on our present state of knowledge.

\begin{itemize}
    \item{\em Network Level Evaluation for LTE-V Systems.} Most of the previous results focus on lower layers performance evaluation, while the network level or above is in general difficult to evaluate due to the highly dynamic network topology and diversified traffic profiles. Although \cite{liu2016coordinative} proposes to link the traffic generator with the network simulator for a complete view, this combined solution fails to catch the physical layer adaptation for LTE-V systems, such as the V2V channel modeling and modified frame structure. 
    \item{\em End-to-End Throughput/Delay Profile for Platooning Systems.} Another issue that is associated with the previous problem is the lack of end-to-end throughput/delay profile characterization for platooning scenarios. Without that information, we are unable to identify the performance bottleneck among different network layers, either in terms of throughput or delay.
    \item{\em Optimized Configuration of Platooning Systems.} In addition, previous results for platooning systems focus on physical layer or MAC layers aspects \cite{zhang2017novel}, while the modeling and evaluation on the end-to-end network layer behavior is limited. Therefore, we shall build the network simulator for V2V communication before we can determine key parameters in the vehicle platooning system, such as the length of car platooning team and the safety distances among different cars.
\end{itemize}

To address the above issues, we start with an integrated platform that combines traffic generator and network simulator together, and build the V2V transmission capability according to LTE-V specification \cite{3gpp:36.885} correspondingly. Based on that, we simulate the end-to-end throughput and delay profiles in different layers to compare different configurations of platooning systems. Through numerical experiments, we show that the LTE-V system is unable to support the highest degree of automation under shadowing effects in the vehicle platooning scenarios, which requires ultra-reliable low-latency communication enhancement in 5G networks. Meanwhile, the throughput and delay performance for vehicle platooning changes dramatically in PDCP layer, where we believe further improvements are necessary. 

The rest of this paper is organized as follows. A summary of V2V transmission and the corresponding combined simulator is elaborated in Section~\ref{sect:bg}. In Section~\ref{sect:ps}, we build the end-to-end LTE-V based simulator for platooning systems by modifying the channel model, the frame structure as well as the performance evaluator. Numerical results are given in Section~\ref{sect:sr} and concluding remarks are summarized in Section~\ref{sect:conc}.

\section{Preliminaries} \label{sect:bg}
In this section, we provide a summary of platooning and LTE-V systems, and introduce the corresponding combined end-to-end simulator in what follows.

\subsection{Summary of Platooning and LTE-V} \label{spl}
Vehicle platooning, by allowing vehicles to travel together, is shown to maximize highway throughput, reduce the traffic drug, and improve driving safety and comfort levels of automated driving simultaneously \cite{ploeg2014lp}. In order to support vehicle platooning capability, on board unit (OBU) with direct V2V communication ability to interchange important parameters in automated longitudinal control systems will be compulsory. In addition, as vehicle platooning systems often operate at a relatively high speed and the inter-vehicle distances need to keep small for traffic efficiency, a high reliable and low latency communication protocol is therefore required. 

Due to the similarity between device-to-device (D2D) and V2V communication,  the sidelink mode 1 and mode 2, previously supporting D2D communication, have been upgraded to mode 3 and mode 4 for V2V communication in LTE R14, where mode 3 focuses on network supported vehicular communication and mode 4 targets to provide V2V direct transmission capability. With vehicle platooning scenario, we mainly consider the V2V communication in this paper and summarize the main features of the sidelink mode 4 in the following.

\begin{itemize}
    \item{\em Higher Layer} One of the major differences between the sidelink mode 2 and mode 4 transmissions is the potential wireless resources. In mode 4, to support purely distributed scheduling with bi-directional transmission capability, higher layer parameters such as constant bit rate (CBR) measurement and resource pool configuration need to be reconfigured for lower layer processing.
    \item{\em MAC Layer} Different from the collision avoidance mechanism in DSRC, which monitors the current channel status in real time, LTE-V use sensing based semi-persistent scheduling (SPS) in MAC layer. It detects channel status in last 1000 ms to estimate the current channel occupancy to guarantee reliability. Resource reservation is maintained by informing other transmitters in sidelink control information (SCI) to further reduce collisions.
    \item{\em Physical Layer} To support the high Doppler effect, enhanced demodulation reference signal (DMRS) scheme is adopted, where the number of reference signals per frame is increased from two to three or four symbols. In addition, to reduce the collision probability, wireless resources are divided into sub-channel basis with fixed physical sidelink control channel (PSCCH) size. We refer interested readers to \cite{3gpp:frame} for more detailed explanation.
\end{itemize}

\begin{figure}[h] \centering 
\includegraphics[width= 3.3in]{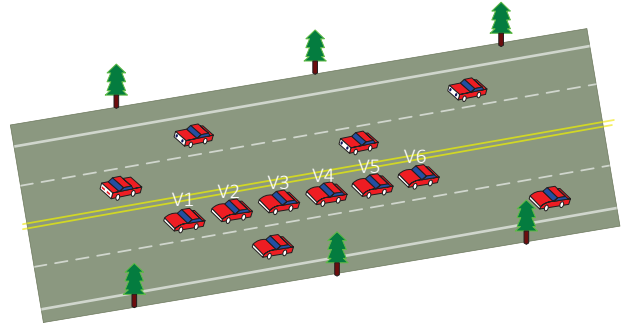}  %{vehicle-platoon.jpg} 
\caption{An illustrative example of vehicle platooning system with the platooning length equal to six. With vehicle platooning, the head car of platooning $V_1$ are broadcasting control messages to the remaining cars through wireless links. In this scenario, the messages have to be timely and reliably delivered to all the remaining cars for safety reasons.} 
\label{platoon}
\end{figure}

\subsection{Combined Simulator}
With the highly dynamic network topology and diversified traffic profile, traditional network simulators, such as network simulator version 3 (NS-3) \cite{NS3.org} or objective modular network testbed in C++ (OMNeT++) \cite{omnet.org}, are in general difficult to support. To solve this issue, one feasible solution is to combine  network simulator with traffic generator, e.g. simulation of urban mobility (SUMO) as reported in \cite{behrisch2011sumo}. Through this approach,  traffic generator can offer a variety of configurable parameters to model a customized simulation environment and smoothly transfer to the network simulator through traceExporter interfaces. Typical examples of the combined generator include ``SUMO + NS-3'' for neighbor selection \cite{liu2016coordinative} or ``SUMO + OMNeT++'' for SPS performance evaluation in V2V communication \cite{molina2017system}.

However, to support network level evaluation for LTE-V based vehicle platooning systems, especially when the channel imperfectness and the frame structure modification are taken into consideration, a higher layer modification as proposed in \cite{molina2017system} will {\em not} be sufficient.

\section{End-to-End LTE-V based Network Simulator} \label{sect:ps}
\begin{figure}[h]
\centering  
\includegraphics[height=11cm,width=8cm]{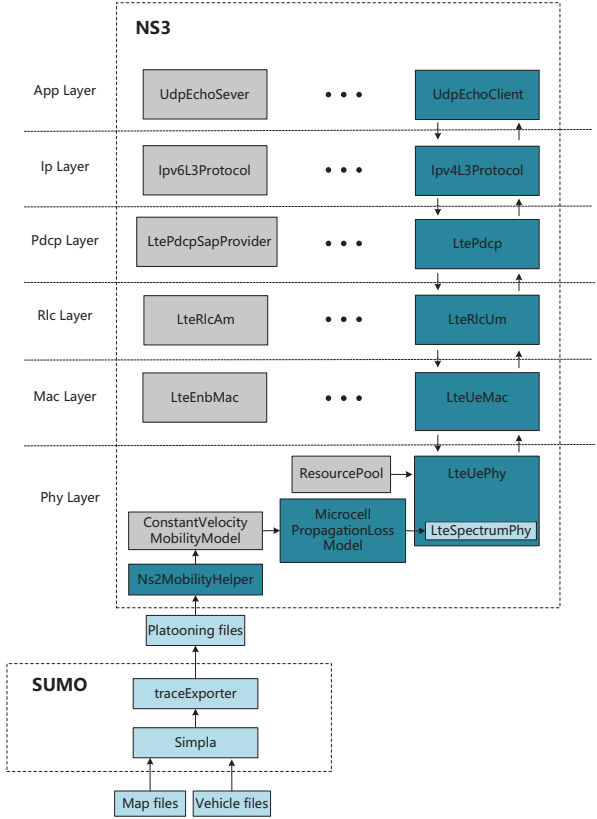}  \caption{An overview of proposed modifications in the combined simulator. To be more specific, we modify modules including Ns2MobilityHelper, MicrocellPropagationLossModel, LteUePhy and LteUeMac to support mobility model, platooning channel model, and LTE-V frame structure respectively. LteUeMac, LteRlcUm, LtePdcp, Ipv4L3Protocol and UdpEchoClient are modified to estimate the throughput and delay profiles for each layer.}
\label{modification}  
\end{figure}

In this section, to address the aforementioned design challenges in LTE-V based vehicle platooning systems, based on the D2D implementation in NS-3 \cite{rouil2017implementation}, we specifically adjust the channel modeling, the frame structure and the performance evaluation parts as shown in Fig.~\ref{modification}.

\subsection{Platooning Channel Modeling}

Comparing with the conventional D2D communication channel model as implemented in \cite{rouil2017implementation}, the shadowing effects of V2V communication are no longer stable due to the high mobility of vehicles. In this sense, LTE-V has proposed to use a block-wise shadowing model, where the shadowing coefficient remains static within the 100ms block and varies according the following equation in a block-by-block manner \cite{3gpp:36.885}. Mathematically, the shadowing component in the $n$-th block, $S_n$, is given by, 
\begin{eqnarray}
S_n = e^{-d_n/d_{cor}} \times S_{n-1}-\sqrt{1-e^{-2d_n/d_{cor}}} \times N_n,
\label{shad} 
\end{eqnarray}
where $S_1$ is generated from a log-normal distribution with standard deviation equals to 3dB. $d_n$ denote the moving distance in the $n$-th block and $d_{cor}$ denote the decorrelation distance respectively. $N_n$ is the independent coefficient, which is also generated from a log-normal distribution with standard deviation equals to 3dB. In the platooning scenario, each vehicle has different shadowing effect according to \eqref{shad}, which may lead to different conclusion from D2D case. As a result, we modify the MicrocellPropagationLossModel module in the NS-3 platform as shown in Fig.~\ref{modification}, which is able to model the pathloss as well as shadowing effects in the V2V platooning systems.

\subsection{Frame Structure Adjustment} \label{sect:fsa}

\begin{figure}[h] \centering 
\includegraphics[height=4.5cm,width=9cm]{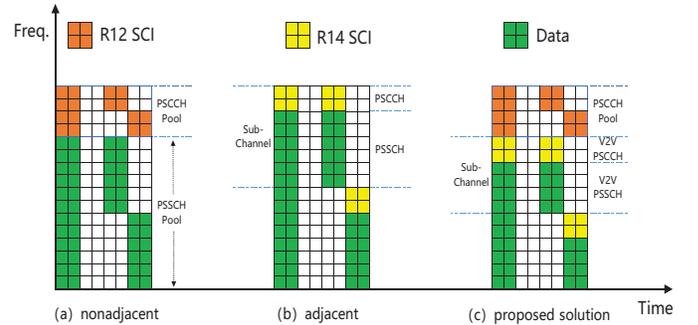}  
\caption{Wireless resource allocation for LTE R12 sidelink (D2D, non-adjacent allocation), R14 sidelink (V2V, adjacent allocation), and proposed solution. In D2D and V2V configurations, the PSCCH and PSSCH are allocated in a non-adjacent/adjacent approach respectively, while in the proposed scheme, we combine these two schemes together to minimize the potential interference to the NS-3 platform as explained in Section~\ref{sect:fsa}.} 
\label{frame}
\end{figure}

According to LTE-V specification \cite{3gpp:frame}, the sidelink channel configuration has been updated to support adjacent resource allocation and sub-channel based resource allocation as shown in Fig.~\ref{frame} (b), where the physical sidelink control channel (PSCCH) and the physical sidelink shared channel (PSSCH) occupy the resources in an adjacent manner\footnote{Note that it allows to aggregate different sub-channels for high volume data transmission as well.}. In the NS-3 platform, to directly change the frame structure and operating mechanism is quite challenging due to the following reasons. First, the sidelink control information allocation and decoding philosophy is quite different, e.g. the D2D based (R12) frame structure has pre-determined PSCCH location and the LTE-V based (R14) scheme provides per sub-channel based PSCCH assignment as shown in Fig.~\ref{frame}. Second, with distributed resource allocation requirement for V2V transmission, disjoint resource combing will be necessary and the corresponding PSCCH decoding mechanism needs to be significantly updated.  

To overcome the above issues\footnote{The original design priciple for NS3 platform is using centralized control for all the communication entities. Although LTE-V based V2V communication relies on a distributed implementation scheme, we still need some centralized control processes, such as registration and synchronization, to make the V2V system work properly.}, we proposed to use a hybrid solution, where we keep the traditional PSCCH pool as usual and manipulate the LTE-V based scheme purely on the previous PSSCH pool as shown in Fig.~\ref{frame}(c). Meanwhile, we further allow the traditional PSCCH pool to decode R12 sidelink control information (SCI) successfully in order to
minimize the potential influence to V2V transmission. Based on this proposed scheme, we modify LteUePhy and LteUeMac modules in the NS-3 platform accordingly.

\subsection{Performance Evaluator}

\begin{figure}[h]
\centering  
\includegraphics[height=7cm,width=9cm]{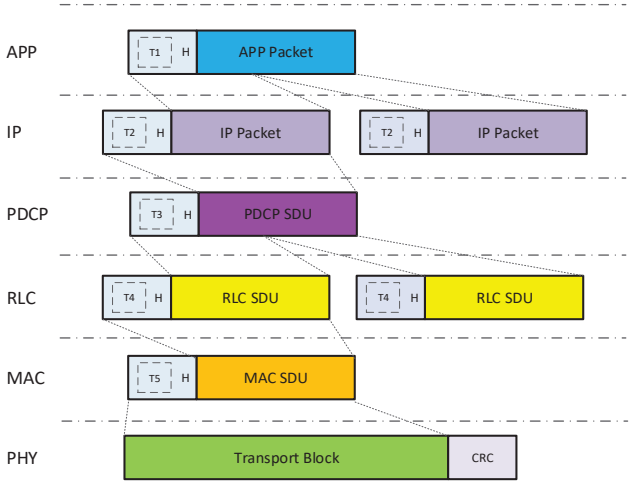}  
\caption{We added some tags (T1-T5), which contain the sending time, to the headers of the packages in each layer. By calculating the difference between the sending time and receiving time in the receiving side, end-to-end delay of each layer is obtained.}
\label{eva-latency}
\end{figure}

In addition, in order to characterize the delay/throughput profiles of V2V networks, we propose to add Tags in head of packets of each layer as shown in Fig.~\ref{eva-latency},  and add monitoring modules in each layers. By calculating the differences between the sender and receiver as well as the successfully received information bits, we can generate the delay/throughput profiles directly. In the implementation, we modify each layer (UdpEchoClient, Ipv4L3Protocol, LtePdcp, LteRlcUm and LteUeMac) to generate head and add the monitoring modules as shown in Fig.~\ref{modification}.

\section{Simulation Results} \label{sect:sr}

In this section, extensive numerical results are presented for vehicle platooning systems using LTE-V transmission protocol. Based on the proposed combined simulator, we give several examples to answer the following three questions. {\em 1) What is the suitable transmission scheme for safety and infotainment messages? 2) What is the maximum platooning length of the vehicle communication environment? 3) Which layer is the most critical in terms of throughput and delay?} The simulation environment is based on Fig.~\ref{platoon} and detailed parameters are listed in Table~\ref{para}.

\begin{table} [h] 
\centering 
\caption{Simulation Parameters for Vehicle Platooning under Shadowing Effects}  
\label{para} 
\footnotesize
\begin{tabular}{c c | c c}  
\toprule
Parameter & Value & Parameter & Value \\

\midrule
Distribution &  Log-normal  & 
Number of vehicles & 9  \\

\midrule  
Std. Deviation &  3 dB &
Inter-vehicle Distance  & 2-5 m \\

\midrule 
$d_{cor}$ &  25 m &
 Packet size baseline & 72 bytes \\

\midrule 
Antenna height  & 1.5 m &
Sending interval baseline & 20 ms \\

\midrule 
Noise power  & -116 dBm &
Safety message size & 20 bytes \\

\midrule 
Average speed  & 50.4 km/h &
Safety message interval  & 10 ms \\

\midrule 
Simulation time  & 45 s &
 &  \\

\bottomrule
\end{tabular}  
\end{table}

\subsection{Effects of Different Applications} 
\label{diffapp}

\begin{figure}[h]
\centering  
\includegraphics[height=5cm,width=7cm]{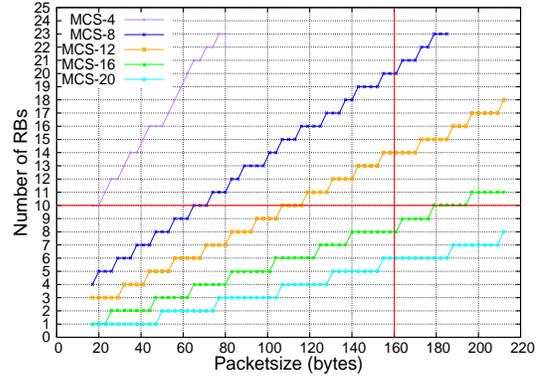}  \caption{The number of RBS change with packet sizes under the sending interval baseline for different MCS.}
\label{packet}  
\end{figure}

\begin{figure}[h]
\centering  
\includegraphics[height=5cm,width=7cm]{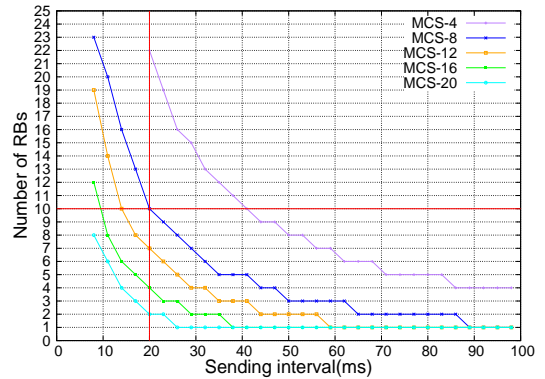}  
\caption{The number of RBs change with the transmitting interval under the packet size baseline for different MCS.}
\label{interval}  
\end{figure}

As packet size and packet transmission interval change in the application layer, the physical layer shall select the appropriate MCS and number of RBs to achieve the required throughput. Intuitively, we need to increase the number of RBs and the MCS level to satisfy certain throughput requirements. Therefore in this section, we take the packet size and packet transmission interval as two indicators to evaluate the throughput requirements. As we can see from Fig.~\ref{packet}, the number of RBs grows monotonically with the packet size while the sending interval is below the baseline shown in Table~\ref{para}. In addition, when the MCS level increases, the slope will be smaller. Under the packet size baseline in Table~\ref{para}, a tight delay (sending interval in this case) often requires more RB resources as shown in Fig.~\ref{interval}. As a summary, for safety messages (e.g. delay is less than $20$ ms and occupied RB is less than $10$), MCS level $12$ or above should be supported; while for infotainment information (e.g. packet size is more than $160$ bytes and occupied RB is less than $10$), MCS level $16$ or above should be supported.

\subsection{Platooning length}
\label{platlen}

\begin{figure}[h]
\centering  
\includegraphics[height=7cm,width=7cm]{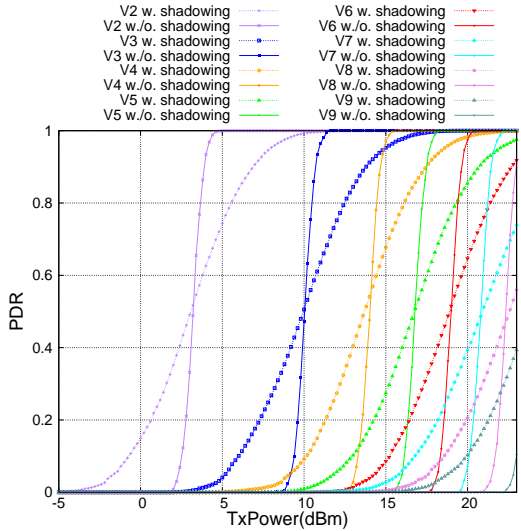} 
\caption{Platooning length under MCS:4 and number of RBs:24 with/without shadowing. V1-V9 represent the vehicle in the vehicle platooning system shown in Fig.~\ref{platoon}.}
\label{4-24} 
\end{figure}

\begin{figure}[h]
\centering  
\includegraphics[height=7cm,width=7cm]{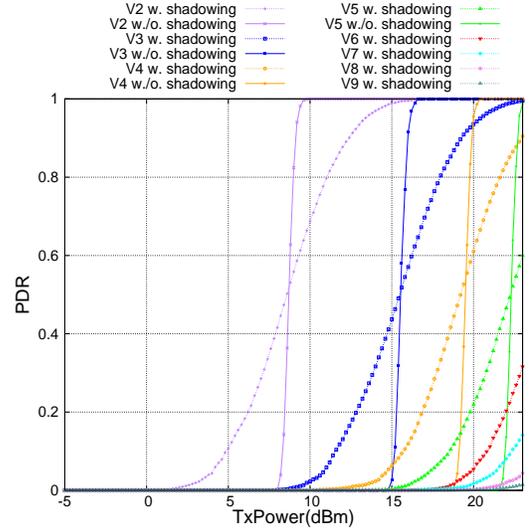} 
\caption{Platooning length under MCS:20 and number of RBs:4 with/without shadowing. V1-V9 represent the vehicle in the vehicle platooning system shown in Fig.~\ref{platoon}. }
\label{20-4} 
\end{figure}

From the above simulation, we can see that for determined application layer packet size and packet interval, different combinations of MCS and RB numbers can be selected to satisfy the throughput requirement. However, different MCS and RB numbers require different channel conditions, and the maximum number of vehicles in the platooning need to be adjusted as well. Here we investigate the performance of transmitting safety message, the parameter of which is given in Table~\ref{para} in the platooning system as shown in Fig.~\ref{platoon}. Specifically, we compare the packet delivery rate (PDR, defined as the ratio between the number of received packets and the number of transmitted packets.) curves for different vehicles in the platooning system under different shadowing conditions in Fig.~\ref{4-24} and Fig.~\ref{20-4}, where the dashed and solid lines represent the PDR under shadowing \& non-shadowing cases respectively.

Both in Fig.~\ref{4-24} and Fig.~\ref{20-4}, it can be seen that the gap between curves of each vehicle gradually decreases. This is because there is a logarithmic relationship between pathloss and distance. With the same application layer packet size and transmitting interval, it can be seen that the scheme of smaller MCS and more RB will bring larger platooning length. 

On the other hand, the platooning length under non-shadowing cases is always larger than shadowing cases. In shadowing case, we can see the slope is smaller, especially in the range between $0.9$ and $1$, which results in most of the receiving vehicles hovering between $0.9$ and $1$ and therefore can not reach high reliability.

As shown in TABLE~\ref{platoon-compare}, we compare our best simulation result (MCS level $4$, and $24$ required RBs) with the proposed platooning length under different reliability requirements based on R15 \cite{3gpp:platoon}. L1-L5 represent the autonomous driving level, which is given by Society of Automotive Engineers(SAE)  \cite{sae2014taxonomy}. We can see that the platooning length is always satisfied under L1-L2. Under L3-L5, the platooning length is enough under non-shadowing cases but insufficient under shadowing cases.

\begin{table} [h] 
\centering 
\caption{The requirements specified in the standard and comparison with our results}  
\label{platoon-compare}
\footnotesize
\begin{tabular}{c c c}  
\toprule
autonomous driving level & L1-L2 & L3-L5 \\

\midrule
platooning length &  $\geq$ 5  &  $\geq$ 5  \\

\midrule  
latency &  $\leq$ 25 ms  & $\leq$ 10 ms \\

\midrule  
reliability &  $\geq$ 90\%  & $\geq$ 99.99\% \\
\midrule 
our experiments w./o. shadowing  &
7  & 6 \\

\midrule 
our experiments w. shadowing  &
5  & 2 \\

\bottomrule
\end{tabular}  
\end{table}

\subsection{End-to-End Throughput/Delay Profile}
\label{endpro}

\begin{figure}[h]
\centering  
\includegraphics[height=6cm,width=8cm]{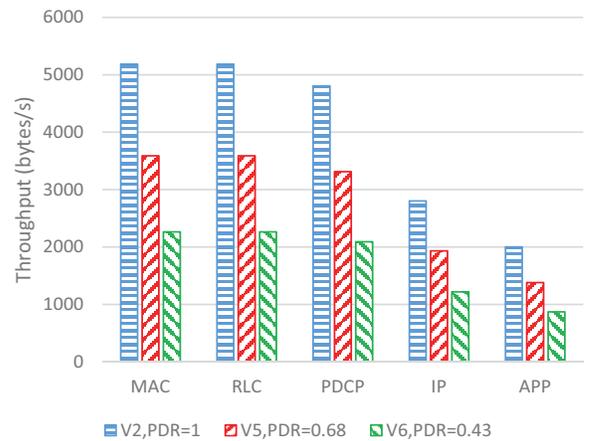} 
\caption{Throughput of each layer. Three sets of experimental results are given. V2, V5, and V6 represent the second vehicle, the fifth vehicle, and the sixth vehicle, respectively. Their PDR is given in the figure.}
\label{sta-thr} 
\end{figure}

\begin{figure}[h]
\centering  
\includegraphics[height=6cm,width=8cm]{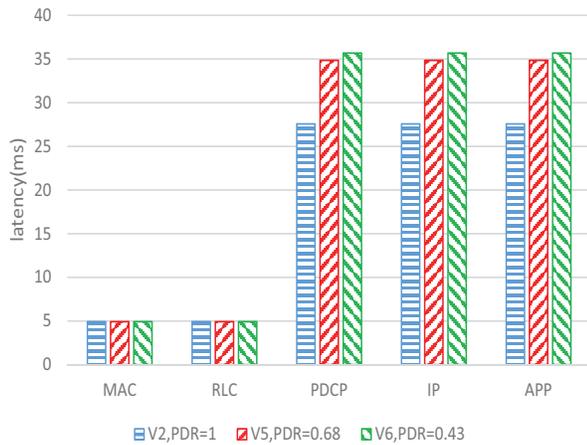} 
\caption{Delay of each layer, Three sets of experimental results are given. V2, V5, and V6 represent the second vehicle, the fifth vehicle, and the sixth vehicle, respectively. Their PDR is given in the figure.}
\label{sta-lat} 
\end{figure}

We show the throughput and delay profile of the second (V2), fifth (V5) and sixth (V6) vehicles in the platooning, which is obtained from our monitoring module, in Fig.~\ref{sta-thr} and Fig.~\ref{sta-lat}. In Fig.~\ref{sta-thr}, we can see that the main increase in throughput cost occurs between the IP and PDCP layers. In Fig.~\ref{sta-lat}, we can see that it spends a lot of time between the RLC layer and the PDCP layer, which we have to optimize to satisfy the delay requirement. Further, when PDR decreases, the delay will increase; we believe it is caused by reordering timer in RLC Layer, which is designed to detect loss of RLC PDUs at lower layer \cite{3gpp:rrc}.

\section{Conclusion} \label{sect:conc}

In this paper, we provide a network level simulator for LTE-V based V2V platooning systems. On top of the combined simulation environment generated from NS-3 and SUMO, we modify the building modules across different layers to support V2V communication. Through numerical results, we find that LTE-V specification can not support vehicle platooning requirements as defined in \cite{3gpp:platoon}, where ultra-reliable low-latency communication enhancement for 5G networks is necessary. Meanwhile, based on the investigation of end-to-end throughput/delay profile for different layers, we argue that PDCP may limit the overall system performance for V2V communication, which requires further research efforts.

\section*{Acknowledgement}

This work was supported by the National Key Research and Development Plan of China under No. YS2017YFGH000872, the National Natural Science Foundation of China (NSFC) Grants under No. 61701293, the National Science and Technology Major Project Grants under No. 2018ZX03001009, the Huawei Innovation Research Program (HIRP), and research funds from Shanghai Institute for Advanced Communication and Data Science (SICS).

\bibliographystyle{IEEEtran}
\bibliography{IEEEfull,ref}

% If references do not show as normal, please change the compiler to LaTeX. yutao

\end{document}